\documentclass[12pt]{article}
\usepackage{hyperref}
\usepackage{tabularx}
\usepackage{graphicx}
\usepackage{multirow}
\usepackage{booktabs}
\usepackage{array}
\usepackage{booktabs} 
\usepackage{booktabs,subcaption,amsfonts,dcolumn}
\usepackage{soul}
\usepackage{authblk}
\usepackage{natbib}
\usepackage{url}
\usepackage{amsmath}
\newcommand{\blind}{0}
\usepackage{enumitem}
\usepackage[doublespacing]{setspace}
\usepackage{bm}
\usepackage{amssymb}
\usepackage[font=scriptsize]{caption}

\DeclareMathOperator{\Cov}{Cov}
\DeclareMathOperator{\E}{E}

\DeclareMathOperator*{\argmin}{argmin}

\newcolumntype{M}[1]{>{\centering\arraybackslash}m{#1}}

\makeatletter
\newcommand*{\addFileDependency}[1]{
  \typeout{(#1)}
  \@addtofilelist{#1}
  \IfFileExists{#1}{}{\typeout{No file #1.}}
}
\makeatother

\newcommand*{\myexternaldocument}[1]{%
    \externaldocument{#1}%
    \addFileDependency{#1.tex}%
    \addFileDependency{#1.aux}%
}

\usepackage{xr}
\myexternaldocument{supplementary}

\newcommand{\specialcell}[2][c]{%
  \begin{tabular}[#1]{@{}c@{}}#2\end{tabular}}

\begin{document}

\def\spacingset#1{\renewcommand{\baselinestretch}%
{#1}\small\normalsize} \spacingset{1}


\if0\blind
{
  \title{An Adaptive Multivariate Functional EWMA Control Chart}
   
\author[1]{Christian Capezza}
\author[2]{Giovanna Capizzi}
\author[1]{Fabio Centofanti}
\author[1*]{Antonio Lepore}
\author[1]{Biagio Palumbo}
 
\affil[1]{Department of Industrial Engineering, University of Naples Federico II, Piazzale Tecchio 80, 80125, Napoli, Italy}
\affil[2]{Department of Statistical Sciences, University of Padua, Via Cesare Battisti 241, 35121, Padova, Italy}
\affil[*]{Corresponding author: antonio.lepore@unina.it}

\renewcommand\Affilfont{\itshape\small}
\date{}
\maketitle

\large{\textbf{This is an original manuscript of an article published by Taylor \&
Francis in \textit{Journal of Quality Technology} on 10 October 2024, available at: \href{https://doi.org/10.1080/00224065.2024.2383674}{https://doi.org/10.1080/00224065.2024.2383674}}}

} \fi

\if1\blind
{
  \bigskip
  \bigskip
  \bigskip
  \begin{center}
    {\LARGE\bm Title}
\end{center}
  \medskip
} \fi

\bigskip
\begin{abstract}
In many modern industrial scenarios, the measurements of the quality characteristics of interest are often required to be represented as functional data or profiles.
This motivates the growing interest in extending traditional univariate statistical process monitoring (SPM) schemes to the functional data setting.
This article proposes a new SPM scheme, which is referred to as adaptive multivariate functional EWMA (AMFEWMA), to extend the well-known exponentially weighted moving average (EWMA) control chart from the univariate scalar to the multivariate functional setting.
The favorable performance of the AMFEWMA control chart over existing methods is assessed via an extensive Monte Carlo simulation. 
Its practical applicability is demonstrated through a case study in the monitoring of the quality of a resistance spot welding process in the automotive industry through the online observations of dynamic resistance curves,  which are associated with multiple spot welds on the same car body and recognized as the full technological signature of the process.
\end{abstract}

\noindent%

\vfill
\newpage

\spacingset{1.45}

\section{Introduction}
\label{sec:intro}
The main aim of statistical process monitoring (SPM) is to detect special causes of variation possibly acting on a  process, which is then said to be out of control (OC).
Otherwise,  it is said to be in control (IC).
As is known, SPM is currently implemented in two phases.
The first one (Phase I) is concerned with identifying a clean data set to be assumed as representative of the IC state of the process, named Phase I sample, while the second one (Phase II) aims at the prospective monitoring of new observations \citep{qiu2013introduction}, often referred to as Phase II observations or data set.

In modern industrial SPM, the capability of new data acquisition systems is reshaping the size and the variety of signals and measurements available, which are often functions of time or space and can be best characterized by functional variables or profiles \citep{ramsay,kokoszka}.
The simplest approach for the monitoring of one or multiple quality characteristics in this form is the application of classical SPM techniques for multivariate data \citep{montgomery2020introduction} on scalar features extracted from them, although this simple approach risks being problem-specific, arbitrary, and flattening useful information.

To avoid that, profile monitoring \citep{noorossana2011statistical} aims to directly monitor quality characteristics in the form of one or multiple functional variables. 
Some recent examples  are those of  \cite{jin1999feature,colosimo2010comparison,zou2012lasso,chou2014monitoring,grasso2014profile,paynabar2016change,grasso2016using,grasso2017phase,menafoglio2018profile,maleki2018overview,wang2018thresholded,ren2019phase,jones2021practitioners,centofanti2020functional,capezza2021functional,capezza2022robust}.
These contributions are based on functional extensions of the Shewart control chart and allow
the process state to be assessed only on the information acquired at the current point in time, without considering previous data points, which may contain decisive information about process condition trends.
For this reason, the Shewart control chart may be particularly ineffective in detecting small persistent shifts.

The exponentially weighted moving average (EWMA) control chart is instead known to overcome this limitation by evaluating the current state of the process through a statistic calculated as a weighted average of the observation at the current point in time and the statistic itself at the previous point. 
The weighted average is determined by a single weighting parameter, usually denoted with $\lambda$ \citep{qiu2013introduction}, such that $0 \leq \lambda \leq 1$. 
Although the design, implementation, and properties of the EWMA control chart have been widely discussed for both univariate and multivariate quality characteristics \citep{crowder1989design,gan1993optimal,serel2009economic,chan2000some,lucas1990exponentially,capizzi2003adaptive,knoth2007accurate,lowry1992multivariate, jones2001performance}, 
few extensions have been proposed in the profile monitoring literature in this regard.
It is worth mentioning \cite{fasso2016functional}, who developed a functional EWMA  control chart for both   univariate and multivariate functional data.
In particular, they applied a multivariate EWMA control chart to the basis coefficients resulting from a chosen finite-dimensional representation of the functional data in the reference sample.
\cite{ren2019phase} move forward with the integration of the EWMA scheme
with the multichannel functional principal component analysis, developed by \cite{paynabar2016change} for the Phase II monitoring of multichannel profiles.

The major challenge for any EWMA control chart is related to the choice  of the  weighting parameter $\lambda$, which should minimize the average run length (ARL) defined as the average number of observations required to signal an OC state with respect to a specified shift \citep{lucas1990exponentially}.
The ARL corresponding to no shift is usually referred to as IC ARL.   
It is known that the smaller the value of $\lambda$, the smaller the shift with respect to which the ARL is the minimum for a given IC ARL. 
However, in most real cases, there is no prior information available on the magnitude of the shift that may actually occur in the OC scenario. 

Therefore, methods are needed for adaptively selecting $\lambda$ such that the resulting EWMA control chart performs adequately in a wide range of OC conditions.
In the univariate scalar setting, this issue is addressed by the adaptive EWMA (AEWMA) control chart proposed by \cite{capizzi2003adaptive} that smoothly combines EWMA and Shewhart control charting schemes, by adaptively selecting the weighting parameter through a suitable function of the magnitude of the difference between the current observation and the  monitoring statistic observed up to the previous point in time. 
This idea has been extended to the multivariate setting \citep{mahmoud2010multivariate,haq2019adaptive}, but has never been addressed in the profile monitoring literature.

In this work, we propose a new method for the monitoring of a process through observations of a multivariate functional quality characteristic, which is named adaptive multivariate functional EWMA (AMFEWMA) control chart and extends the underlying idea of \cite{capizzi2003adaptive} to the multivariate profile monitoring  setting. 
As a case study, we consider the monitoring of a resistance spot welding (RSW) process in the automotive body-in-white manufacturing  \citep{zhang2011resistance}. RSW  is commonly used to join overlapping steel galvanized sheets by means of two copper electrodes to guarantee the structural integrity and solidity of welded assemblies in each vehicle \citep{martin2014assessment}. 
In this context, the dynamic resistance curve (DRC) is considered as highly representative of the physical and metallurgical development of a spot weld \citep{capezza2021functional_clustering}, and therefore, of the quality of the joint produced.
In particular, there is a consensus among experts regarding the correlation between the latter and the degradation status or wear level of the electrodes \citep{manladan2017review}. 
A control chart capable of accommodating varying degrees of mean shift, contingent upon different wear levels, without prior knowledge about the current wear level, can prove immensely advantageous.

The paper is structured as follows. Section \ref{sec_meth} introduces the AMFEWMA control chart. 
Section \ref{sec_simulation} reports a Monte Carlo simulation study of the performance of the AMFEWMA control chart in identifying process mean shifts of a multivariate functional quality characteristic with respect to other non-adaptive SPM approaches that have already been presented in the literature.
The practical applicability of the proposed method is
demonstrated in Section \ref{sec_real} through a case study in the SPM of multivariate DRCs from a RSW process in automotive body-in-white manufacturing.
Section \ref{sec_con} concludes the paper.
Supplementary Materials that provide details on data generation in the simulation study are available online.
All computations and plots have been obtained using the R programming
language \citep{r2021}.

\section{The AMFEWMA control chart}
\label{sec_meth}

The elements of the AMFEWMA control chart are developed in the following sections.
Section \ref{sec_smoothing} shows how smooth functional data are obtained from the original discrete observations for each curve;
Section \ref{sec_mfcpa} illustrates the multivariate functional principal component analysis (MFPCA); Section \ref{sec_monitoring} introduces the AMFEWMA scheme, while its design and implementation details are referred to Section \ref{design}.

\subsection{Data Smoothing}
\label{sec_smoothing}

For each profile, measurements $y_i$, $i = 1, \dots, m$, of the quality characteristic are supposed to be collected in a discrete fashion with the observation points $t_1, \dots, t_m \in \mathcal T \subset \mathbb R$, into a realization $X$ of a random function with values in $L^2(\mathcal T)$, i.e., the Hilbert space of square integrable functions defined on the compact set $\mathcal{T} \subset \mathbb R$, with inner product between $a,b \in L^2(\mathcal T)$ given by $ \langle a, b \rangle = \int_{\mathcal T} a (t) b (t) dt$. 
Hence, methods are required to convert discrete raw data $\lbrace (t_i, y_i) \rbrace_{i=1,\dots,m}$ into functional data $X$.
If the discrete data are assumed without any measurement error, functional data can be theoretically drawn up by merely connecting the whole set of points $\lbrace (t_i, y_i) \rbrace$. 
However, this does not represent the ordinary situation. When measurement error is present, each discrete observation is expressed as
\begin{equation}
\label{eq_1}
     y_i=X\left(t_i\right)+\varepsilon_{i}, \quad i=1,\dots, m, 
\end{equation}
for $i=1,\dots,m$, where $\varepsilon_{i}$ are independent identically distributed random errors with zero mean.
The functional variable $X$ is intrinsically infinite dimensional, that is  infinite features would be needed to completely specify it at each possible argument  $t\in \mathcal{T}$.
Therefore, from Equation \eqref{eq_1}, data smoothing techniques are commonly used to recover  $X$ by discarding exogenous perturbation due to error terms $\varepsilon_{i}$.
A common approach consists of representing $X$ through a linear combination of $K$ known basis functions $\bm{\phi}=\left( \phi_1,\dots,\phi_K\right)^T$.
That is,
\begin{equation}
     X\left(t\right)=\sum_{k=1}^{K}c_{k}\phi_k\left(t\right)=\bm{c}^T\bm{\phi}\left(t\right) \quad t\in\mathcal{T},
\end{equation}
where $\bm{c}=\left(c_{1},\dots,c_{K}\right)^T$ is the vector of basis coefficients.
The value of $K$  is not crucial unless it is sufficiently large to capture the local behavior of the functional data \citep{cardot2003spline}.
Then, the problem of recovering $X$ from the discrete raw data reduces to finding the estimate $\hat{\bm{c}}$ of the unknown coefficient vector $\bm{c}$, which is obtained by minimizing  the following  penalized sum of squares error 
\begin{equation}
\label{eq_opt}
    \hat{\bm{c}}=\argmin_{\bm{c}\in\mathbb{R}^K}\sum_{i=1}^m\left(y_i-\bm{c}^T\bm{\phi}\left(t_i\right)\right)^2+\lambda_s \bm{c}^T\bm{R} \bm{c},
\end{equation}
where $\lambda_s>0$ is a smoothing parameter and $\bm{R} $ is a matrix whose entries are $\langle \phi^{\left(d\right)}_i\left(t\right), \phi^{\left(d\right)}_j \rangle$, with $\phi^{\left(d\right)}$ denoting the $d$-th derivative of $\phi$.
The smoothing parameter $\lambda_s$ is chosen by minimizing the generalized cross-validation (GCV) criterion, which is a well-known method to achieve a trade-off between variance and bias. This criterion takes into account the degrees of freedom of the estimated curve that vary according to $\lambda_s$. 
The reader may want to refer to \cite{ramsay} for further details.
The matrix $\bm R$, which controls the penalty on the right-hand side of Equation \eqref{eq_opt}, is usually computed by setting  $d=2$, i.e.,  by penalizing the function roughness.

The functional datum $\hat X(t)$ so estimated is
\begin{equation}
\label{eq_4}
    \hat{X}\left(t\right)=\hat{\bm{c}}^T\bm{\phi}\left(t\right) \quad t\in\mathcal{T}.
\end{equation}
The most common choice for $\bm{\phi}$ in case of non-periodic functional data is the B-spline basis system, owing good computational properties and great flexibility \citep{ramsay}. 
More in general, splines are known to be optimal in the sense of being the smoothest functions interpolating the data \citep{green1993nonparametric}. Spline functions divide the functional domain into subintervals, by means of breakpoints. Over any subinterval, the spline is a polynomial of order $q$, with $q-1$ non-zero derivatives and matching proper derivative constraints between adjacent polynomials \citep{de1978practical}.

\subsection{Multivariate Functional Principal Component Analysis}
\label{sec_mfcpa}
Let us consider a random vector $\bm X = (X_1, \dots, X_p)^T$ with realizations in $\mathbb{H}$, the Hilbert space of $p$-dimensional vectors of functions in $L^{2}(\mathcal{T})$.
Given two elements ${\bm a}, {\bm b} \in \mathbb H$, where $\bm a = (a_1, \dots, a_p)^T$ and $\bm b = (b_1, \dots, b_p)^T$, the inner product of $\mathbb{H}$ is defined as $\langle {\bm a}, {\bm b}\rangle_{\mathbb H} = \sum_{j=1}^p \langle a_j, b_j \rangle$.
We assume that ${\bm X}$ has mean $\bm{\mu}=\left(\mu_1,\dots,\mu_p\right)^T$, where $\mu_j(t)=\E(X_{j}(t))$, $j = 1,\dots, p$, $t\in\mathcal{T}$ and covariance $\bm{C}=\lbrace C_{jk}\rbrace_{1\leq j,k \leq p}$, $C_{jk}(s,t)=\Cov(X_{j}(s),X_{k}(t))$, $s,t\in \mathcal{T}$.
From the multivariate Karhunen-Lo\`{e}ve's Theorem \citep{happ2018multivariate} it follows that
\begin{equation*}
	\bm{X}(t) - \bm \mu (t) = \sum_{l=1}^{\infty} \xi_{l}\bm{\psi}_l(t),\quad t\in\mathcal{T},
\end{equation*}
where $\xi_{l}=\langle \bm{\psi}_l, \bm{Z}\rangle_{\mathbb{H}} $ are random variables, say \textit{principal components scores} or simply \textit{scores}, such that  $\E\left( \xi_{l}\right)=0$ and $\E\left(\xi_{l} \xi_{m}\right)=\rho_{l}\delta_{lm}$, with $\delta_{lm}$ denoting the Kronecker delta.
The elements of the orthonormal set $\lbrace \bm{\psi}_l\rbrace_{l\geq 1}$, $\bm{\psi}_l=\left(\psi_{l1},\dots,\psi_{lp}\right)^T$, with $\langle \bm{\psi}_l,\bm{\psi}_m\rangle_{\mathbb{H}}=\delta_{lm}$, are referred to as \textit{principal components}, and are the  eigenfunctions  of the covariance $\bm{C}$  of $\bm{Z}$ corresponding to the eigenvalues $\rho_1\geq\rho_2\geq \dots\geq 0$.

MFPCA aims to estimate $\bm \psi_l$ and $\rho_l$, $l \geq 1$, starting from $N$ independent realizations, $\bm{X}_1, \dots, \bm{X}_N$, of $\bm{X}$, where $\bm X_i = (X_{i1}, \dots, X_{ip})^T$, which can be obtained through the data smoothing technique described in Section \ref{sec_smoothing} and denoted by $\hat{\bm{X}}_{1}, \dots, \hat{\bm{X}}_{N}$.
That is, each $X_{ij}$, $j=1,\dots,p$, is estimated as $\hat X_{ij} (t) = \hat{\bm c}_{ij}^T \bm \phi (t)$, where $ \hat{\bm c}_{ij} = (\hat c_{ij1}, \dots, \hat c_{ijK})^T$. 
Following the basis function expansion approach of \cite{ramsay}, each  component $\psi_{lj}$ of $\bm \psi_l$ is represented as the following linear combination of the $K$ basis functions $\phi_{1},\dots,\phi_{K}$ used to obtain $\hat{\bm{X}}_{1}, \dots, \hat{\bm{X}}_{N}$
\begin{equation}
	\label{eq_appcov}
	\psi_{lj}(t)= \sum_{k=1}^{K} b_{ljk}\phi_{k}(t), \quad  t\in\mathcal{T}, \quad j=1,\dots,p, \quad l=1,2,\dots,
\end{equation}
where $\bm{b}_{lj}=\left(b_{lj1},\dots,b_{ljK}\right)^T$ are the eigenfunction coefficient vectors. 
Under these assumptions, MFPCA \citep{ramsay,chiou2014multivariate} consists of performing standard multivariate principal component analysis of the random vectors $\bm{W}^{1/2} (\hat{\bm{c}}_{i} - \overline{\bm c})$, where $\hat{\bm{c}}_{i}=\left(\hat{\bm{c}}_{i1}^{T},\dots,\hat{\bm{c}}_{ip}^{T}\right)^T$, $\overline{\bm c} = \sum_{i=1}^N \hat{\bm{c}}_{i} / N$ and $\bm{W}$ is a block-diagonal matrix with diagonal blocks  $\bm{W}_j$, $j=1,\dots,p$, whose entries are $w_{k_1 k_2} = \langle\phi_{k_1},\phi_{k_2}\rangle$, $k_1,k_2=1,\dots,K$. 
Specifically, the estimated eigenvalues $\hat{\rho}_{l}$ of $\bm{C}$ are the eigenvalues  of the sample covariance matrix $\bm{S}$ obtained from $\bm{W}^{1/2} (\hat{\bm{c}}_{i} - \overline{\bm c})$, whereas
$\hat{\bm{\psi}}_{l}=\left(\hat{\psi}_{l1},\dots,\hat{\psi}_{lp}\right)^{T}$  is obtained through Equation \eqref{eq_appcov}, where $\bm{b}_{lj}=\bm{W}^{-1/2}\bm{u}_{lj}$ and $\bm{u}_{l}=\left(\bm{u}_{l1}^T,\dots,\bm{u}_{lp}^T\right)^T$ is the $l$-th eigenvector of $\bm{S}$. 

In practice, it is assumed that, as the eigenvalues decrease toward zero, the leading eigenfunctions tend to reflect the most important features of $\bm{X}$. 
That is, $\hat{\bm{X}}_{i}$ are approximated by $\hat{\bm{X}}_{i}^L$ through the following truncated principal component decomposition
\begin{equation}
	\label{eq_appx}
	\hat{\bm{X}}_{i}^L(t)= \hat{\bm{\mu}}_{}(t)+\hat{\bm{D}}_{}(t)\sum_{l=1}^{L} \hat{\xi}_{il}\hat{\bm{\psi}}_{l}(t) \quad t\in\mathcal{T},
\end{equation}
where $\hat{\bm{D}}_{}$ is a $p \times p$ diagonal matrix whose diagonal entries are $\hat{v}_{j}^{1/2}$ and $\hat{\xi}_{il}= \langle \hat{\bm{\psi}}_{l}, \hat{\bm{X}}_{i}\rangle_{\mathbb{H}}$.
In the profile monitoring literature, the parameter $L$ is generally chosen such that the retained principal components explain at least a given percentage of the total variability, which is usually in the range 70-90$\%$ \citep{paynabar2016change,ren2019phase,centofanti2022real,capezza2022robust}, even though more sophisticated methods can be used as well \citep{jolliffe2016principal,capezza2020control}.

\subsection{The monitoring scheme}
\label{sec_monitoring}

Let us consider a sequence of independent realizations in $\mathbb H$ of the random vector $\bm X_n = (X_{n1}, \dots, X_{np})^T$, $n=1,2,\dots$.
Suppose that (a) before an unknown observation, say $n_{0}$, the process is in control and the mean is equal to a target value $\bm{\mu }_{0}$, (b) after $n_{0}$, the mean becomes $\bm{\mu}_{1}$ = $\bm{\mu}_{0}$ + $\bm{\delta}$, where $\bm{\delta}$ is an unknown shift, and (c) we want to detect the occurrence of this shift as soon as possible.
Without loss of generality, it is assumed that, when the process is in control, the data are mean centered, i.e., $\bm \mu_0$ is a vector of zero functions.
As in \cite{fasso2016functional}, an EWMA control chart for multivariate functional data can be built based on the statistic $\tilde {\bm Y}_{n} = (\tilde Y_{n1},\dots,\tilde Y_{np})^T$, defined as 
\begin{equation}
\label{eq_MFEWMA}
         \tilde{\bm Y}_{n}(t) = 
         ({\bm I} - {\bm\Lambda})\tilde{\bm Y}_{n-1}(t) + {\bm \Lambda}  {\bm X}_{n}(t) =
         \tilde{\bm Y}_{n-1}(t) + {\bm\Lambda} \tilde {\bm E}_n (t)
         ,\quad t \in \mathcal{T},n=1,2,\dots,
     \end{equation}
where $\tilde {\bm E}_n (t) = (\tilde E_{n1} (t),\dots, \tilde E_{np} (t))^T = {\bm X}_{n} (t) - \tilde{\bm Y}_{n-1} (t) $, ${\bm I}$ is the $p\times p$ identity matrix and ${\bm \Lambda}=\text{diag}(\lambda_{1},\lambda_{2},...,\lambda_{p})$, with $\lambda_{j} \in (0,1]$, $j=1,\dots,p$, the weighting parameters. 
For $n=1$, $\tilde{\bm Y}_0$ represents the starting value, often set equal to the target value $\bm \mu_0$.

By elaborating on Equation \eqref{eq_MFEWMA}, we propose the AMFEWMA statistic ${\bm Y}_{n} = (Y_{n1},\dots,Y_{np})^T$, defined as follows
\begin{equation}
     \label{AMFEWMA_statistic}
         {\bm Y}_{n}(t) = 
         ({\bm I} - {\bm\Lambda}_{n}(t)){\bm Y}_{n-1}(t) + {\bm \Lambda}_{n}(t)  {\bm X}_{n}(t)=
         {\bm Y}_{n-1}(t) + {\bm \Lambda}_{n} (t) \bm{E}_{n} (t),
         \quad t \in \mathcal{T},n=1,2,\dots.
     \end{equation}
Similarly to $\tilde{\bm E}_n(t)$, $ {\bm E}_n (t) = ( E_{n1} (t),\dots,  E_{np} (t))^T = {\bm X}_{n}(t) - {\bm Y}_{n-1} (t)$; while, differently from $\bm \Lambda$, ${\bm \Lambda}_{n} (t) = \text{diag}(w(E_{n1} (t)),\dots,w(E_{np} (t)))$, where
$w(E_{nj} (t)) = \eta(E_{nj} (t))/E_{nj} (t)$ and $\eta(E_{nj} (t))$ is a score function evaluated at the error $E_{nj} (t)$, $j=1,\dots,p$. 
In this way, the AMFEWMA statistic ${\bm Y}_{n}$ results in a weighted average of the current observation ${\bm X}_{n}$ and the charting statistic value at the previous time point ${\bm Y}_{n-1}$, as in a conventional EWMA chart, but with weights $w(E_{nj}(t))$ changing over time. 
That is, the score functions $\eta$ to be used in Equation \eqref{AMFEWMA_statistic} are proposed by extending the score functions for a univariate scalar quality characteristic, proposed by \cite{capizzi2003adaptive}, to the multivariate functional data setting.
A score function $\eta(\cdot)$  must be strictly increasing, odd ($\eta(x) = \eta(-x)$), and such that $\eta(E_{nj}(t))$ is close to $\lambda E_{nj}(t)$ (resp. to $E_{nj}(t)$) when $|E_{nj}(t)|$ is small (resp. is large).
Based on these properties, we propose to use as the score function either
\begin{center}
\begin{equation}
\label{eq_score1}
    \eta_{1}(E_{nj}(t)) = 
    \begin{cases}
    E_{nj}(t)+(1-\lambda) C_j(t) & \text{if } E_{nj}(t) < -C_j(t)\\
    \lambda E_{nj}(t) & \text{if } | E_{nj}(t)|\leq C_j(t)\\
    E_{nj}(t)-(1-\lambda) C_j(t) & \text{if } E_{nj}(t)>C_j(t)
\end{cases}, \quad t \in \mathcal T,
\end{equation}
\end{center}
or
\begin{center}
\begin{equation}
\label{eq_score2}
    \eta_{2}(E_{nj}(t)) = 
    \begin{cases}
    E_{nj}(t)[1-(1-\lambda)(1-(E_{nj}(t)/C_j(t))^2)^2] & \text{if } |E_{nj}(t)|\leq C_j(t)\\
    E_{nj}(t) & \text{otherwise}
\end{cases}, \quad t \in \mathcal T,
\end{equation}
    
\end{center}
where $\lambda \in (0,1]$ and $C_j(t) = k \sigma_{j}(t)$, with $k$ denoting a real positive constant and $\bm \sigma = (\sigma_1, \dots, \sigma_p)^T$ representing the standard deviation function of the current observation $\bm X_n$. 
From Equation \eqref{AMFEWMA_statistic}, it is worth noting that, when ${\bm X}_{n}$ approaches ${\bm Y}_{n-1}$, i.e.,  $E_{nj}(t)$ tends to zero through a given score function, the AMFEWMA chart is expected to perform like the MFEWMA control chart based on the statistic $\tilde{\bm Y}_{n}$ defined in Equation \eqref{eq_MFEWMA},. 
Instead, it is expected to behave as a Shewhart control chart when $|E_{nj}(t)|$ is large.
Hence, the calculation of the AMFEWMA statistic requires the careful selection of the parameters $\lambda$ and $k$, which will be addressed in Section \ref{design}.
The functional extension of the classical Hotelling's statistic, based on the AMFEWMA statistic can be defined as 
\begin{equation}
\label{V2_statistic}
    V_{n}^{2} = \sum_{i=1}^{p} \sum_{j=1}^{p} \int_{\mathcal T} \int_{\mathcal T} Y_{ni}(s) K_{ij}^{*}(s,t) Y_{nj}(t) ds dt,
\end{equation}
where the function $K^*_{ij}$ is defined as
\begin{equation}
\label{inverse_cov}
    K_{ij}^{*}(s,t) = \sum _{l=1}^{L} \frac{1}{\rho_{l}} \psi_{li}(s) \psi_{lj}(t), \quad s,t \in \mathcal T, i,j=1,\dots,p,
\end{equation}   
and ${\bm \psi}_{l} = (\psi_{l1},\dots,\psi_{lp})^{T}$ denotes the eigenfunctions corresponding to the eigenvalues $\rho_{l}$ of the asymptotic covariance $\bm C(s,t)$ of ${\bm Y}_{n}$, calculated by applying the MFPCA presented in Section \ref{sec_mfcpa}. The number of principal components $L$ is usually chosen such that the retained principal components explain a given fraction of the total variability.

\subsection{Design and implementation}
\label{design}
The AMFEWMA control chart will be designed to signal when the statistic 
$V_n^2$ given in Equation \eqref{V2_statistic} exceeds the upper control limit $h$ chosen as the smallest value achieving an  IC ARL larger than or equal to a pre-specified value, usually denoted by $ARL_0$.
However, to calculate the control limit $h$, we need to design the AMFEWMA statistics through the selection of the parameters $\bm{\theta}=(\lambda,k)^T$ involved in the score functions suggested in   Equation \eqref{eq_score1} and Equation \eqref{eq_score2}.
In this paper, the optimal parameters are obtained by elaborating on the idea of \cite{capizzi2003adaptive} to provide, to the best possible extent, the AMFEWMA control chart with good performance at different shift magnitudes.

First, a desired $ARL_0$, together with two shift values, say $\bm \delta_{1}$ and $\bm \delta_{2}$, with $\lVert \bm \delta_{1} \rVert \ll \lVert \bm \delta_{2} \rVert$, and a small positive constant $\varepsilon$ (e.g., $\varepsilon=0.05$) must be chosen. 
In general $\bm \delta_1$ and $\bm \delta_2$ can represent any ``small'' and ``large'' split, respectively, as in \cite{capizzi2003adaptive}.
Then,  the optimal parameters $\bm{\theta}$, denoted by $\bm{\theta}^*$, are found  as the solutions of the following problem
\begin{align}
    \bm \theta_2 &= \argmin_{\bm \theta } ARL(\bm \delta_2, \bm \theta)\quad \text{s. t.} \quad ARL(0, \bm \theta) = ARL_0,\\
    \bm{ \theta}^*  &= \argmin_{\bm \theta } ARL(\bm \delta_1, \bm \theta)\quad \text{s. t.} \quad ARL(0, \bm \theta) = ARL_0 \quad \text{and} \quad ARL(\bm \delta_2, \bm \theta) \leq (1+\varepsilon)ARL(\bm \delta_{2}, \bm \theta_2),
    \label{eq_optimal_theta}
\end{align}
where $ARL(\bm \delta, \bm \theta)$ denotes the ARL achieved by the AMFEWMA control chart with parameters $\bm \theta$ when the mean shift is equal to $\bm \delta$.
The constant $  \varepsilon$ allows the AMFEWMA scheme to achieve an ARL at $\bm{\delta}_2$ that is close to the minimum $ARL(\bm \delta_2, \bm \theta_2)$ while searching for the minimum $ARL(\bm \delta_1, \bm \theta)$ with respect to $\bm \theta$.
Hereinafter, we will denote the AMFEWMA control chart with parameters $\bm \theta^*$ chosen with the above procedure as $\text{AMFEWMA}^*$.
It is worth highlighting that the parameters included in the vector $\bm \theta$, and thus $\bm \theta^*$, may vary with respect to the score function used. 

To calculate the control limit $h$,
$n_{seq}$ sequences of $n_{obs}$ observations are generated through a bootstrap procedure as proposed by \cite{gandy2013guaranteed} from the Phase I sample. 
On each sequence, the AMFEWMA statistic and the function $K^*_{ij}$ are calculated as defined in Equation \eqref{AMFEWMA_statistic} and Equation \eqref{inverse_cov}.
Then, the monitoring statistic $V_{n}^{2}$ is calculated as defined in Equation \eqref{V2_statistic}, respectively. 
For each sequence, the number of observations acquired up to the first signal of an OC state is stored as a run length (RL), then the ARL is calculated as the average of the RL values over the $n_{seq}$ sequences.
The control limit $h$ is then chosen to reach a pre-specified $ARL_0$.

Overfitting issues may arise when calculating $h$ \citep{kruger2012statistical} because the asymptotic covariance of the AMFEWMA statistic $\bm Y_n$, used in Equation \eqref{V2_statistic} to calculate $V_{n}^{2}$, depends on $\bm Y_n$ itself.
Thus, to mitigate this problem, the control limit $h$ will be customarily based on a random subset of the Phase I sample, referred to as the \textit{tuning} set, which is separate from the remaining one used to estimate the covariance of ${\bm Y}_{n}$, referred to as the \textit{training} set.

In the Phase II monitoring, given the current sequence ${\bm X}_{n}$, the AMFEWMA statistic ${\bm Y}_{n}$ and the corresponding monitoring statistic ${V}_{n}^2$ are obtained from Equation \eqref{AMFEWMA_statistic} and Equation \eqref{V2_statistic}, respectively, according to the asymptotic covariance function $\bm C(s, t)$ estimated in Phase I. 
An alarm signal is issued if ${V}_{n}^2$ exceeds the control limit $h$ calculated in Phase I. 
A similar procedure is used in Phase II, where $n^{II}_{seq}$ bootstrap sequences are generated from the Phase II data set, and the ARL is obtained as the average of the RL values calculated for each sequence.

\section{Simulation study}
\label{sec_simulation}
The performance of the proposed AMFEWMA control chart in identifying mean shifts in multivariate functional data is evaluated by means of an extensive Monte Carlo simulation study. 
The data generation process, which is detailed in Supplementary Material A, is performed as in \cite{capezza2022robust} and is inspired by the typical functional form of DRCs in a RSW process, as that introduced in Section \ref{sec:intro} and studied in Section \ref{sec_real}. 
Without loss of generality, multivariate functional quality data are generated with $p = 5$ components. 
Two OC scenarios are considered where the OC observations simulate typical outlying DRCs, based on the work of \cite{xia2019online}. 
In Scenario 1, OC observations mimic a splash weld, which is also known as the expulsion phenomenon and is caused by an excessive welding current flowing through the electrodes. In Scenario 2, they mimic a phase shift of the peak time caused by an overlay large electrode force. 
In each scenario, six increasing severity levels $ SL\in\lbrace 0,1,2,3,4,5,6\rbrace $ are explored, where $SL=0$ means that the process is IC.
The severity level quantifies the extent to which Phase II data set profiles deviate from their Phase I counterparts. 
A higher severity level indicates a greater departure from the baseline and, consequently, a more pronounced OC state for the process.
Supplementary Material A defines the severity levels, which can be visually interpreted in Figure A.1 and A.3 that superimpose, on some IC observations, the shifted mean under each shift type, for severity levels $SL\in\lbrace 1,6\rbrace$.

The proposed chart is compared with two competing approaches: the  EWMA control chart for multivariate  functional data, which is referred to as MFEWMA and based on the Phase II monitoring scheme of \cite{ren2019phase} with weighting parameter $\lambda \in\lbrace 0.1,0.2,0.3,0.5 \rbrace$; and the classical multivariate Shewhart control chart 
 based on the Hotelling $T^2$ statistic applied to the
coefficients obtained from the MFPCA \citep{capezza2023funcharts}, referred to as SHEWHART.

For each scenario and $SL$, 30 simulation runs are performed. Each run considers a Phase I sample composed of 2500 observations, with training and tuning samples of 1000 and 1500 observations, respectively. 
The Phase II data set is composed of 200 sequences of i.i.d. observations with shift locations equal to 100.
The AMFEWMA control chart is implemented as detailed in Section \ref{sec_meth} with $n_{seq}$ and $n_{obs}$ set equal to 500 and 300, respectively. 
The parameter $L$ of Equation \eqref{inverse_cov} is chosen such that the retained principal components explain 90\% of the total variability.

The AMFEWMA and the competing methods' performance is reported in Table \ref{tab_Results_Scenario1} and Table  \ref{tab_Results_Scenario2} in terms of the estimated $ARL$, denoted by $\widehat{ARL}$, as a function of $SL$ for Scenario 1 and Scenario 2, respectively.
When the process is IC, the $\widehat{ARL}$ should be as close as possible to $ARL_0$, which is set equal to 20, while it should be as small as possible when the process is OC.
The AMFEWMA control chart is implemented for $\lambda  \in\lbrace 0.1,0.2,0.3,0.5\rbrace$ and score function $\eta_1(\cdot)$ as reported in Equation \eqref{eq_score1} with $k \in\lbrace 2,3,4\rbrace$, whereas the $\text{AMFEWMA}^*$ is obtained with optimal parameters $\bm \theta^*$  chosen through the algorithm presented in Section \ref{design}. In the latter, the values of   $\bm \delta_1$ and $\bm \delta_2$ are set as $\bm \delta_1= \hat{\bm \mu}_0 + 0.5 \hat{\bm{\sigma}}$ and $\bm \delta_2= \hat{\bm \mu}_0 + 2 \hat{\bm{\sigma}}$, where the sample mean $\bm{\hat{\mu}_0}$ and standard deviation  $\hat{\bm{\sigma}}$ functions are estimated based on 100 bootstrap samples drawn from the Phase I sample.

\begin{figure}
    \centering
        \includegraphics[width=\textwidth]{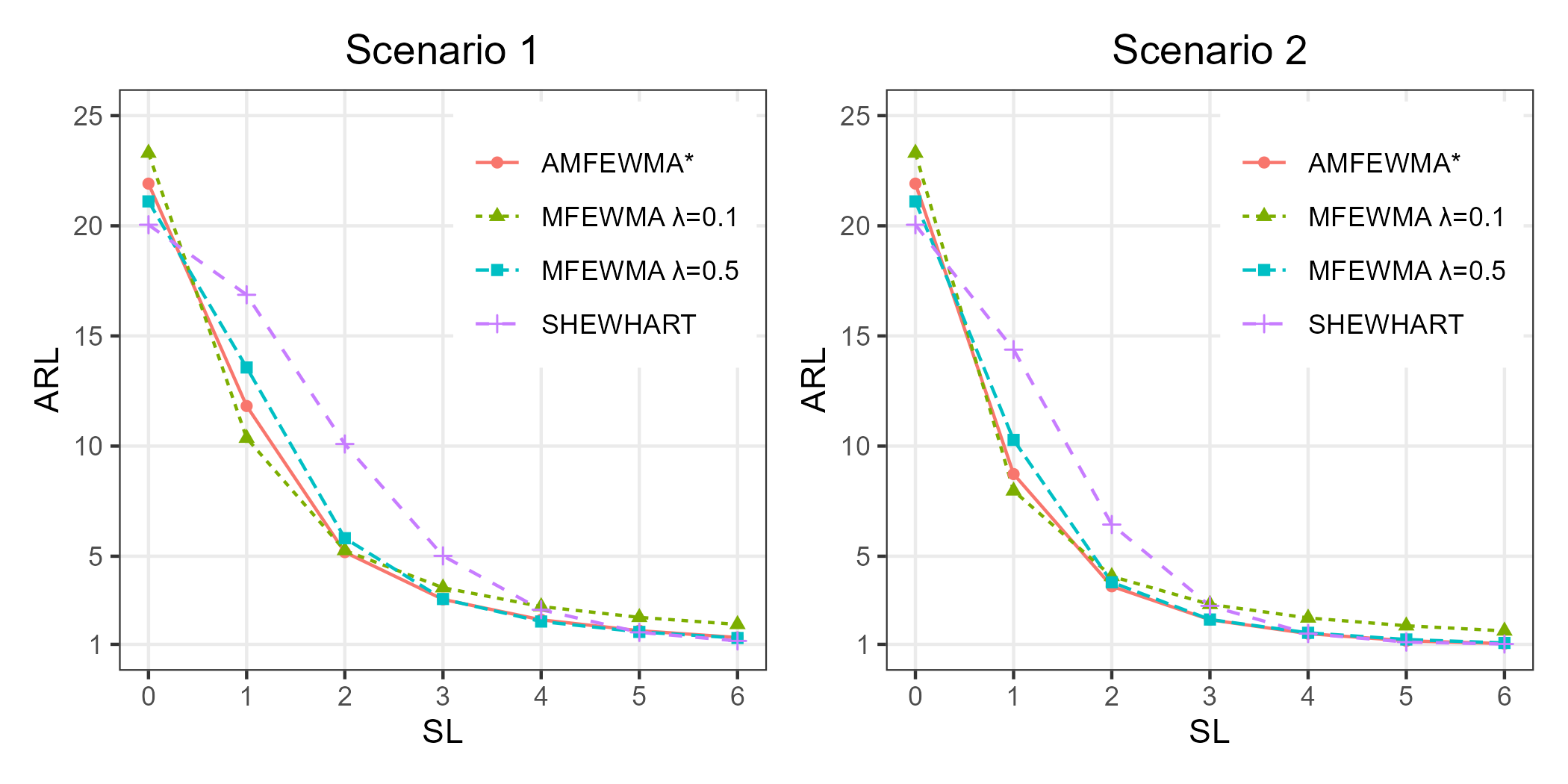}
    \caption{\label{fig:Plot Scenario 1}The estimated  $ARL$ values for the SHEWHART, MFEWMA (with $\lambda \in \lbrace 0.1,0.5\rbrace$) and the $\text{AMFEWMA}^*$ control charts in Scenario 1 and Scenario 2.}   
\end{figure}

In agreement with the literature, Table \ref{tab_Results_Scenario1} and Table \ref{tab_Results_Scenario2} reveal that, among the competing approaches, the SHEWHART control chart achieves the best performance for large  $SL$s, whereas the MFEWMA  with small $\lambda$ values turn out to be the best approach for small $SL$s. 
From these tables, it is clear that the parameters $\lambda$ and $k$  have a marked impact on the performance of the  AMFEWMA control chart in both scenarios.
As an example, for $\lambda = 0.2$ and $k = 2$, the proposed control chart, while being optimal for large shifts, achieves unsatisfactory $\widehat{ARL}$ for small shifts, performing very similarly to SHEWART.
On the other hand, keeping fixed  $\lambda = 0.2$ but setting $k = 4$, the AMFEWMA performs similarly to the MFEWMA. 
In all cases,  for each $SL$, there is a combination of parameters, e.g., $\lambda = 0.3$ and $k = 3$, for which the AMFEWMA performance is the best approach, or is very close to it.
Therefore, the problem of finding optimal parameters $\bm \theta^*$ remains indispensable.
The $\widehat{ARL}$s values achieved by the proposed control chart with optimal parameters $\bm \theta^*$, obtained through the score function $\eta_1$ in Equation \eqref{eq_score1} and referred to as $\text{AMFEWMA}^*$, are reported in the last columns of Table \ref{tab_Results_Scenario1} and Table \ref{tab_Results_Scenario2} and indicate adequate performance for every OC condition considered. 
Indeed, the $\text{AMFEWMA}^*$  performs similarly to the MFEWMA control chart with small $\lambda$ for small $SL$s, and comparably to the SHEWHART control chart for large $SL$s. 
This behavior is  graphically shown in Figure \ref{fig:Plot Scenario 1}, which reports the estimated  $\widehat{ARL}$ values for the SHEWHART, MFEWMA (with $\lambda \in \lbrace 0.1,0.5\rbrace$) and the $\text{AMFEWMA}^*$ control charts in Scenario 1 and Scenario 2.
From these figures,    the $\widehat{ARL}$s  for the $\text{AMFEWMA}^*$ control chart appears always close to the smallest estimated $ARL$ value, for each $SL$  and scenario considered. 

These results are further demonstrated by means of the relative mean index (RMI) suggested by \cite{zou2012lasso} and defined as
\begin{equation*}
    RMI=\frac{1}{6}\sum_{SL=1}^6\frac{ARL_{SL}-sARL_{SL}}{sARL_{SL}},
\end{equation*}
where, at given SL,  $ARL_{SL}$ denotes the estimated $ARL$ value of the considered control chart and $sARL_{SL}$ the smallest estimated $ARL$ value among the competitors. 
The ratio $\frac{ARL_{SL}-sARL_{SL}}{sARL_{SL}}$ can be regarded as a relative efficiency measure of the considered control chart with respect to the best competitor, at given a $SL$, while the $RMI$ represents its average over different SLs. 
The smaller the $RMI$, the better the overall performance of the control chart.
Table \ref{tab:Indicator table} shows the  $RMI$ achieved by the SHEWHART, the MFEWMA (with $\lambda \in \lbrace 0.1,0.2,0.3,0.5\rbrace$), and the $\text{AMFEWMA}^*$ control charts in Scenario 1 and Scenario 2.
In both scenarios, the $\text{AMFEWMA}^*$ confirms to outperform the competing methods also in terms of $RMI$.
It is worth noting that the SHEWHART control chart and MFEWMA scheme with $\lambda=0.1$, which represent practitioner's standard choices, show very poor performance in both Scenario 1 and Scenario 2 and thus they should be used with  care unless the size of the potential OC condition is known a priori.

\begin{table}[ht]
\caption{\label{tab_Results_Scenario1}The estimated $ARL$ values as a function of the severity level $SL$ for Scenario 1. The AMFEWMA control chart is implemented with $\lambda  \in\lbrace 0.1,0.2,0.3,0.5\rbrace$ and $k \in\lbrace 2,3,4\rbrace$. The smallest $\widehat{ARL} $ in each row is highlighted in bold when $SL \neq 0$.}
\centering
\resizebox{1\textwidth}{!}{
\begin{tabular}{ccccccccccccccccccc}
  \toprule
 SL& SHEWHART& \multicolumn{4}{c}{MFEWMA} &\multicolumn{4}{c}{AMFEWMA ($k=2$)} &\multicolumn{4}{c}{AMFEWMA ($k=3$)}&\multicolumn{4}{c}{AMFEWMA ($k=4$)}&$\text{AMFEWMA}^*$ \\\midrule
 & &\multicolumn{4}{c}{$\lambda$} & \multicolumn{4}{c}{$\lambda$} &\multicolumn{4}{c}{$\lambda$} &\multicolumn{4}{c}{$\lambda$}   \\
& &0.1&0.2&0.3&0.5 & 0.1&0.2&0.3&0.5& 0.1&0.2&0.3&0.5& 0.1&0.2&0.3&0.5  \\\midrule
0 & 20.05 & 23.30 & 21.95 & 21.92 & 21.11 & 21.09 & 21.10 & 20.84 & 20.98 & 22.50 & 21.54 & 21.63 & 21.08 & 23.38 & 21.90 & 21.67 & 21.02 & 21.92  \\
1&14.38&\textbf{7.97}&8.04&8.67 &10.28&10.87&10.60&10.75&11.81&8.36&8.29&8.92&10.68&8.00&8.09&8.69&10.35&8.73\\
2&6.44&4.09 & 3.71 &\textbf{3.62} &3.82&4.08&3.86&3.84&4.23&3.81&3.60&3.57&3.84&4.03&3.69&3.60&3.83&3.63\\
3&2.75&2.82 & 2.44 &2.27 &2.13&1.97&\textbf{1.93}&1.93&2.01&2.21&2.14 &2.07&2.05&2.71&2.39&2.23&2.11&2.12\\
4&1.49&2.20 &1.88 &1.72 &1.52& 1.29&\textbf{1.27}&\textbf{1.27}&1.30&1.45&1.47&1.45&1.39&1.98&1.78&1.65& 1.49&1.49\\
5&1.10&1.84 & 1.58&1.42 &1.22&1.06&\textbf{1.05}&1.06&1.06&1.13&1.15 &1.14&1.11&1.50&1.39&1.31&1.18 &1.16\\
6&\textbf{1.01}&1.61 & 1.36 &1.20 & 1.06&\textbf{1.01}&\textbf{1.01}&\textbf{1.01}& 1.01&1.02&1.03&1.03& 1.02&1.19&1.15&1.10&1.04&1.03\\
\bottomrule
   \end{tabular}
}
  \end{table}

\begin{table}[ht]
\caption{\label{tab_Results_Scenario2}The estimated $ARL$ values as a function of the severity level $SL$ for Scenario 2. The AMFEWMA control chart is implemented with $\lambda  \in\lbrace 0.1,0.2,0.3,0.5\rbrace$ and $k \in\lbrace 2,3,4\rbrace$. The smallest $\widehat{ARL}$ in each row is highlighted in bold when $SL \neq 0$.}
\centering
\resizebox{1\textwidth}{!}{
\begin{tabular}{ccccccccccccccccccc}
  \toprule
 SL& SHEWHART& \multicolumn{4}{c}{MFEWMA} &\multicolumn{4}{c}{AMFEWMA ($k=2$)} &\multicolumn{4}{c}{AMFEWMA ($k=3$)}&\multicolumn{4}{c}{AMFEWMA ($k=4$)}&$\text{AMFEWMA}^*$ \\\midrule
 & &\multicolumn{4}{c}{$\lambda$} & \multicolumn{4}{c}{$\lambda$} &\multicolumn{4}{c}{$\lambda$} &\multicolumn{4}{c}{$\lambda$}   \\
& &0.1&0.2&0.3&0.5 & 0.1&0.2&0.3&0.5& 0.1&0.2&0.3&0.5& 0.1&0.2&0.3&0.5  \\\midrule
0 & 20.05 & 23.30 & 21.95 & 21.92 & 21.11 & 21.09 & 21.10 & 20.84 & 20.98 & 22.50 & 21.54 & 21.63 & 21.08 & 23.38 & 21.90 & 21.67 & 21.02 & 21.92  \\
1&16.87&10.36&10.75&11.73 &13.57&13.69&13.49&13.97&14.77&10.94&11.17&12.17&14.04&\textbf{10.38}&10.88&11.79&13.64&11.82\\
2&10.09&5.26 & \textbf{5.04} &5.09 &5.82&6.20&5.95&6.03&6.87&5.24&5.06&5.16&6.03        &5.26&\textbf{5.04}&5.10&5.84&5.19\\
3&5.02&3.57 & 3.17 &3.02 &3.06&3.21&3.08&3.07&3.26&3.29&3.07 &2.97&3.08         &3.53& 3.16&3.02&3.07&3.04\\
4&2.57&2.72 &2.36  &2.19 &2.04& 1.96&1.91&\textbf{1.89}&1.93&2.29&2.16&2.06&1.98         &2.66&2.34&2.17& 2.02&2.11\\
5&1.54&2.22  & 1.91&1.75 &1.56&1.40&1.39&\textbf{1.38}&\textbf{1.38}&1.68&1.63 &1.57&1.48         &2.13&1.86&1.72 &1.55 &1.62\\
6&1.16&1.90 & 1.63&1.49 & 1.29&\textbf{1.13}&\textbf{1.13}&\textbf{1.13}& \textbf{1.13}&1.33&1.31&1.28& 1.20        &1.75&1.56&1.44&1.27&1.31\\
\bottomrule
\end{tabular}
}
  \end{table}
\begin{table}[ht]
\caption{\label{tab:Indicator table}The  $RMI$ for the SHEWHART, MFEWMA (with $\lambda \in \lbrace 0.1,0.2,0.3,0.5\rbrace$), and the $\text{AMFEWMA}^*$ control charts in Scenario 1 and Scenario 2.}
\centering
\resizebox{0.7\textwidth}{!}{
\begin{tabular}{ccccccc}
  \toprule
  & SHEWHART& \multicolumn{4}{c}{MFEWMA} &$\text{AMFEWMA}^*$ \\\midrule
 & &\multicolumn{4}{c}{$\lambda$} &   
 \\
& &0.1&0.2&0.3&0.5 & \\\midrule
Scenario 1 & 1.88 & 2.36 & 1.32 & 0.88 & 0.59 & 0.27 \\ 
Scenario 2 & 2.55 & 1.81 & 0.99 & 0.73 & 0.66 & 0.49 \\ 
   \bottomrule
\end{tabular}
}
\end{table}

\section{Case study}
\label{sec_real}
The case study on the SPM of an RSW process in automotive body-in-white manufacturing, mentioned in the introduction,  is presented to demonstrate the practical applicability of the proposed AMFEWMA control chart as well as to confirm its superior performance over competitors.
Data analyzed are courtesy of Centro Ricerche Fiat and are recorded at the Mirafiori Factory during lab tests on different bodies of the same car model. 
Each body is characterized by the same large number of spot welds and different characteristics, e.g., metal sheet thickness and material and the welding time. 

\begin{figure}
    \centering
        \includegraphics[width=1\columnwidth]{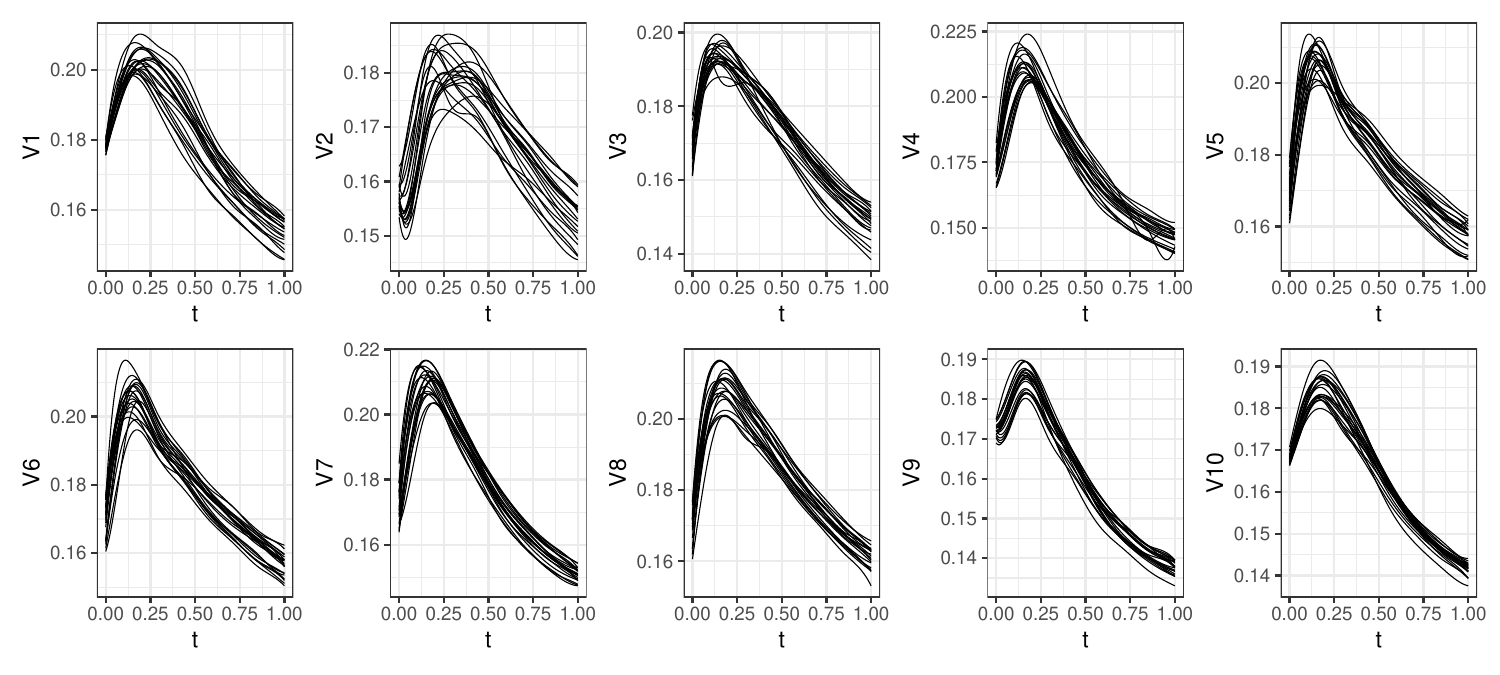}
    \caption{A sketch of DRCs pertaining  to the ten spot welds measured on 20 car bodies randomly selected from the dataset analyzed in the case study.}   
    \label{fig:Real PHASE I data}
\end{figure}

The case study focuses  on a set of $p=10$ spot welds made by the same welding gun on $n=1340$ car bodies.
That is, for each car body, we consider the multivariate functional quality characteristic represented by the vector of $p=10$ DRCs relative to the same ten spot welding points. 
For each DRC, raw measurements were collected at a regular grid of points equally spaced by 1 ms, normalized on the time domain $[0, 1]$.
To provide a sketch of the data, the  10-dimensional vectors of DRCs corresponding to 20 random car bodies out of $n=1340$  are displayed in Figure \ref{fig:Real PHASE I data}.

Even though this is not meant to be the routine application, to allow for further comparison with competing control charting schemes based on this case study, the Phase I sample has been formed by selecting 568 multivariate profiles observed immediately after the renewal, which the electrodes are routinely subject to for neutralizing the wear effect that is known, as mentioned in the introduction, to impact on the quality of the final welded point. 
The remaining 772 observations are grouped into three data sets of 220, 368, and 184 observations, corresponding to three incremental electrode wear levels, namely wear levels 1, 2, and 3, respectively.
These three data sets are then used as Phase II observations to evaluate the proposed chart performance with respect to different levels of shift connected with wear level.   
It is important to emphasize that this is a special case where information regarding electrode wear is available because the data are acquired during lab tests, while such information is commonly out of reach. 
Hence, a control chart capable of accommodating varying degrees of mean shift, contingent upon different levels of wear, without prior knowledge of the wear levels, can prove immensely advantageous.

The AMFEWMA control chart is implemented as in Section \ref{sec_simulation} and the 568 observations of the Phase I sample are randomly split into a training and a tuning set, with  equal sample sizes. 
The optimal parameters $\lambda^* = 0.5$ and $k^* = 4$ used in the $\text{AMFEWMA}^*$ are obtained from Equation \eqref{eq_optimal_theta} through the score function $\eta_1$ reported in Equation \eqref{eq_score1}.
The bootstrap method presented in Section \ref{design} is applied to each of the three Phase II data sets to generate 200 sequences of 200 observations.

\begin{figure}
    \centering
        \includegraphics[width=1\columnwidth]{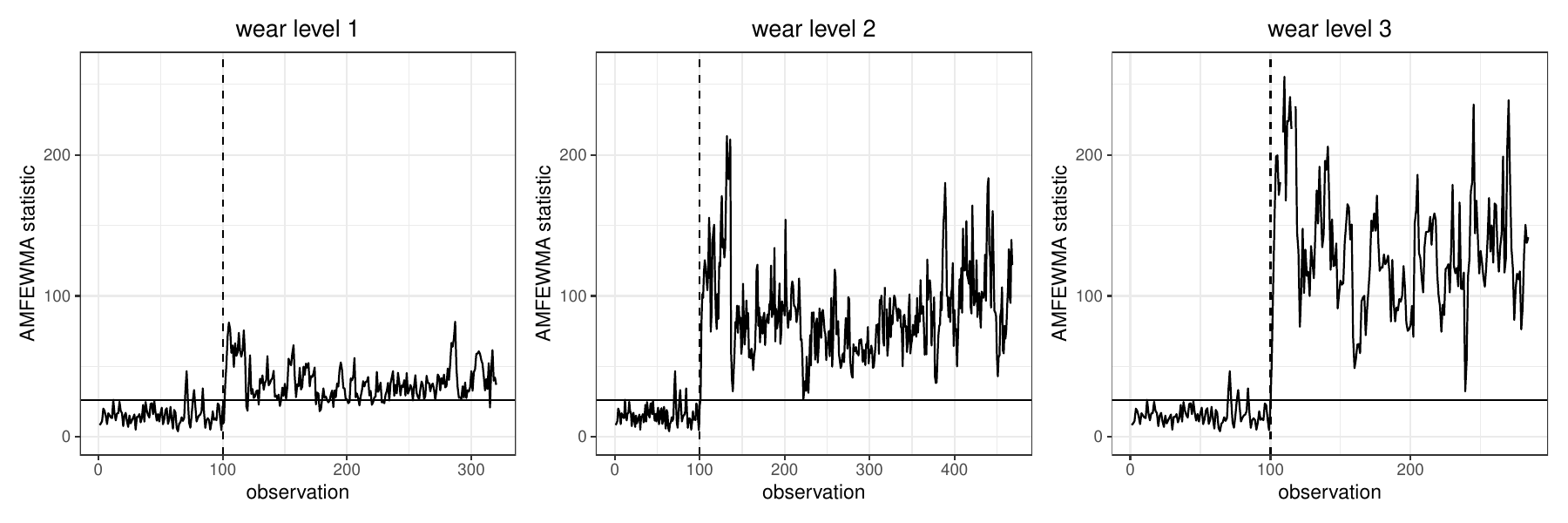}
    \caption{$\text{AMFEWMA}^*$ control chart under wear levels 1, 2, and 3. In each plot, the first 100 observations are randomly sampled from the tuning set, while the Phase II observations are reported after the dashed vertical line.}   
    \label{fig_cc}
\end{figure}

Figure \ref{fig_cc} shows the $\text{AMFEWMA}^*$ application to the Phase II monitoring of the three Phase II data sets corresponding to three increasing wear levels.
In each plot, the first 100 observations are randomly sampled from the tuning set, whereas the Phase II observations for each wear level considered are displayed on the right of the dashed vertical line.
From this figure, the $\text{AMFEWMA}^*$ scheme clearly appears to adequately detect the mean shift under all shift severity levels.
The AMFEWMA performance is further compared in this case study with the competing methods, in terms of ARL, and the results are shown in Table \ref{tab:Tabella risultati dati reali}.
\begin{table} [tb]\small
\caption{\label{tab:Tabella risultati dati reali} ARL calculated on the three Phase II data sets (corresponding to  wear level  1,  2, and  3, respectively) in the case study, for the proposed AMFEWMA and the competing control charts. In bold, the lowest ARL value for each row is reported.}
\centering
\begin{tabular}[t]{c|c|cccc|c}
   & SHEWHART & \multicolumn{4}{c|}{MFEWMA} &  \multirow{2}{*}{\specialcell{$\text{AMFEWMA}^*$\\$\lambda=0.5,k=4$}}\\
   Wear level & & $\lambda$=0.1 & $\lambda$=0.2 & $\lambda$=0.3 & $\lambda$=0.5 & \\
\hline
1 & 5.68 & 3.69 & 3.095 & 2.91 & \textbf{2.905} &  2.92\\
2 & \textbf{1.25} & 2.185 & 1.715 & 1.52 & 1.315 & 1.3\\
3 & \textbf{1.025} & 1.695 & 1.415 & 1.295 & 1.12 & 1.085 \\
\end{tabular}
\end{table}
For wear level 1, the MFEWMA control chart with $\lambda=0.5$ achieves the lowest ARL value, whereas in the other cases, the SHEWHART control chart achieves the best performance. 
It is easy to note that, as expected, the performance of the AMFEWMA chart is always very close to the best-performing competitor.
Therefore,  it can be regarded as the best overall choice in all cases where there is no prior information about the severity of the OC scenario.

\section{Conclusions}
\label{sec_con}
A new control chart for the statistical process control of multivariate functional data is proposed and referred to as adaptive multivariate functional exponentially weighted moving average (EWMA) control chart (AMFEWMA). 
The AMFEWMA chart is the first EWMA scheme that is able to adaptively monitor a multivariate functional quality characteristic. 
Whereas, methods already presented in the literature either apply the conventional EWMA approach to multivariate or univariate functional data. 
Through an extensive Monte
Carlo simulation, the performance of the AMFEWMA chart is compared with two competing methods, which proved the AMFEWMA to outperform competing methods in the correct identification of out-of-control (OC) scenarios when the magnitude of the mean shift is unknown. 
For small shifts, the AMFEWMA performs as the conventional MFEWMA control chart, whereas, for large shifts, it performs as the Shewart control chart. 
The ability of the proposed method to deal with multivariate functional data and effectively signal process mean shifts in a wide range of severity levels is of paramount importance to protect modern industrial processes where many functional quality characteristics are routinely available in line against OC scenarios with different severity levels.

\section*{Supplementary Materials}
The Supplementary Materials contain additional details about the data generation process in the simulation study as well as the R code to reproduce graphs and results over competing methods.

\section*{Data Availability Statement}
The authors confirm that the data supporting the findings of this study are available within the article and its supplementary materials.

\section*{Acknowledgements}

The authors are extremely grateful to CRFWCM R\&I (World Class Manufacturing Research and Innovation) sites of Orbassano and CampusMelfi, for the access to experimental data and the technological insights in the interpretation of the results.

\section*{Funding}

The contribution of A. Lepore and F. Centofanti has been economically supported by Piano Nazionale di Ripresa e Resilienza (PNRR) - Missione 5
Componente 2, Investimento 1.3-D.D. 1551.11-10-2022, PE00000004 within the Extended
Partnership MICS (Made in Italy - Circular and Sustainable).
The activity of B. Palumbo is supported by the MOST - Sustain-
able Mobility National Research Center and received funding from the European Union
Next-GenerationEU (PIANO NAZIONALE DI RIPRESA E RESILIENZA (PNRR) - MIS-
SIONE 4 COMPONENTE 2, INVESTIMENTO 1.4 - D.D. 1033 17/06/2022, CN00000023).
This manuscript reflects only the authors’ views and opinions, neither the European Union
nor the European Commission can be considered responsible for them.

\bibliographystyle{apalike}
\setlength{\bibsep}{5pt plus 0.3ex}
{\small
\bibliography{ref}
}
\end{document}


\def\spacingset#1{\renewcommand{\baselinestretch}
{#1}\small\normalsize} \spacingset{1}

\if0\blind
{
  
\title{Supplementary Materials to ``An Adaptive Multivariate Functional EWMA Control Chart''}
\author[]{}

\setcounter{Maxaffil}{0}
\renewcommand\Affilfont{\itshape\small}
\date{}
\maketitle
} \fi

\spacingset{1.45}
\appendix
\numberwithin{equation}{section}

\section{Details on Data Generation in the Simulation Study}
\label{sec_appA}
The data generation process is inspired by the real-case study in Section 4 and mimics typical behaviors of DRCs in an RSW process.
The data correlation structure is generated similarly to \cite{centofanti2020functional,capezza2022robust}.
The compact domain $\mathcal{T}$ is  set, without loss of generality, equal to $\left[0,1\right]$, and the number components $p$ is set equal to 5.
The eigenfunction set $\lbrace \bm{\psi}_i\rbrace $ is generated by considering the correlation function $\bm{G}$ through the following steps.
\begin{enumerate}
\item Set the diagonal elements $G_{ll}$, $l=1,\dots,p$ of $\bm{G}$ as the \textit{Bessel} correlation function of the first kind \citep{abramowitz1964handbook}. The general form of the correlation function and parameter used  are listed in Table 
Then, calculate the eigenvalues $\lbrace\eta_{lk}^{X}\rbrace$ and the corresponding eigenfunctions $\lbrace\vartheta_{lk}\rbrace$, $k=1,2,\dots$,  of $G_{ll}$, $l=1,\dots,p$.
\item Obtain the cross-correlation function $G_{lj}$, $l,j=1,\dots,p$ and $l\neq j$, by
\begin{equation}
G_{lj}\left(t_1,t_2\right)=\sum_{k=1}^{\infty}\frac{\eta_{k}}{1+|l-j|}\vartheta_{lk}\left(t_{1}\right)\vartheta_{jk}\left(t_{2}\right)\quad t_1,t_2\in\mathcal{T}.
\end{equation}
\item Calculate the eigenvalues $\lbrace\lambda_i\rbrace$ and the corresponding eigenfunctions  $\lbrace \bm{\psi}_i\rbrace $ through the spectral decomposition of $\bm{G}=\lbrace G_{lj}\rbrace_{l,j=1,\dots,p}$, for $i=1,\dots,L^{*}$.
\end{enumerate}
\begin{table}
\caption{Bessel correlation function and parameter for data generation in the simulation study.}
\label{ta_corf}

\centering
\resizebox{0.5\textwidth}{!}{
\begin{tabular}{ccc}
\toprule
&$\rho$&$\nu$\\
\midrule
$J_{v}\left(z\right)=\binom{|z|/\rho}{2}^{\nu}\sum_{j=0}^{\infty}\frac{\left(-\left(|z|/\rho\right)^{2}/4\right)^{j}}{j!\Gamma\left(\nu+j+1\right)}$&0.125&0\\[.35cm]

\bottomrule
\end{tabular}}
\end{table}
Further,  $L^{*}$ is set equal to $10$.
Let   $\bm{Z}=\left(Z_1,\dots,Z_p\right)$ as
\begin{equation}
\bm{Z}=\sum_{i=1}^{L^{*}}\xi_i\bm{\psi}_i.
\end{equation}
with $\bm{\xi}_{L^{*}}=\left(\xi^{X}_1,\dots,\xi^{X}_{L^{*}}\right)^{T}$   generated by means of a  multivariate normal distribution with covariance $\Cov\left(\bm{\xi}_{L^{*}}^{X}\right)=\bm{\Lambda^{X}}=\diag\left(\lambda_1,\dots,\lambda_{L^{*}}\right)$.

Furthermore, let  the mean process $m$  
\begin{multline}
    m(t)= 0.2074 + 0.3117\exp(-371.4t) +0.5284(1 - \exp(0.8217t))\\ -423.3\left[1 + \tanh(-26.15(t+0.1715)) \right]\quad t\in\mathcal{T}.
\end{multline}
Note that the mean function $m$  is generated to resemble a typical DRC through the phenomenological model for the RSW process  presented in \cite{schwab2012improving}.
Then, let define the  contamination models $C_E$ and $C_P$, which mimics a splash weld (expulsion) and phase shift of the peak time,  as 
\begin{equation}
    C_E(t)=\min\Big\lbrace 0, -2M_E(t-0.5)\Big\rbrace \quad t\in\mathcal{T},
\end{equation}
and
\begin{multline}
    C_P(t)= -m(t)-(M_P/20)t + 0.2074\\ + 0.3117\exp(-371.4h(t)) +0.5284(1 - \exp(0.8217h(t)))\\ -423.3\left[1 + \tanh(-26.15(h(t)+0.1715))\right]  \quad t\in\mathcal{T},
\end{multline}
where $h:\mathcal{T}\rightarrow\mathcal{T}$ transforms the temporal dimension $t$ as follows 
\begin{equation}
    h(t)= \begin{cases} 
    t & \text{if } t\leq 0.05 \\
	\frac{0.55-M_P}{0.55}t-(1+\frac{0.55-M_P}{0.55})0.05 & \text{if } 0.05< t\leq 0.6 \\
		\frac{0.4+M_P}{0.4}t+1-\frac{0.4+M_P}{0.4} & \text{if }  t> 0.6, \\
	\end{cases}
\end{equation} 
and $M_E$ and $M_P$ are  contamination sizes.
Then,  the 
 model to generate $\bm{X}=\left(X_1,\dots,X_p\right)^T$ is 
\begin{equation}
\label{eq_modgen}
\bm{X}\left(t\right)=  \bm{m}(t) +\bm{Z}\left(t\right)\sigma+\bm{\varepsilon}\left(t\right) +B_E\bm{C}_E(t)+B_P\bm{C}_P(t) \quad t\in \mathcal{T},
\end{equation}
where $\bm{m}$ is a $p$ dimensional vector with components  equal to $m$, $\sigma>0$, $\bm{\varepsilon}=\left(\varepsilon_1,\dots,\varepsilon_p\right)^T$, where $\varepsilon_i$ are  white noise functions such that for each $ t \in \left[0,1\right] $, $ \varepsilon_i\left(t\right) $ are  normal random varaibles with zero mean and standard deviation $ \sigma_e $,  $\bm{C}_E=\left(C_E,\dots,C_E\right)^T$ and $\bm{C}_P=\left(C_P,\dots,C_P\right)^T$.

\begin{figure}
    \centering
        \includegraphics[width=.9\columnwidth]{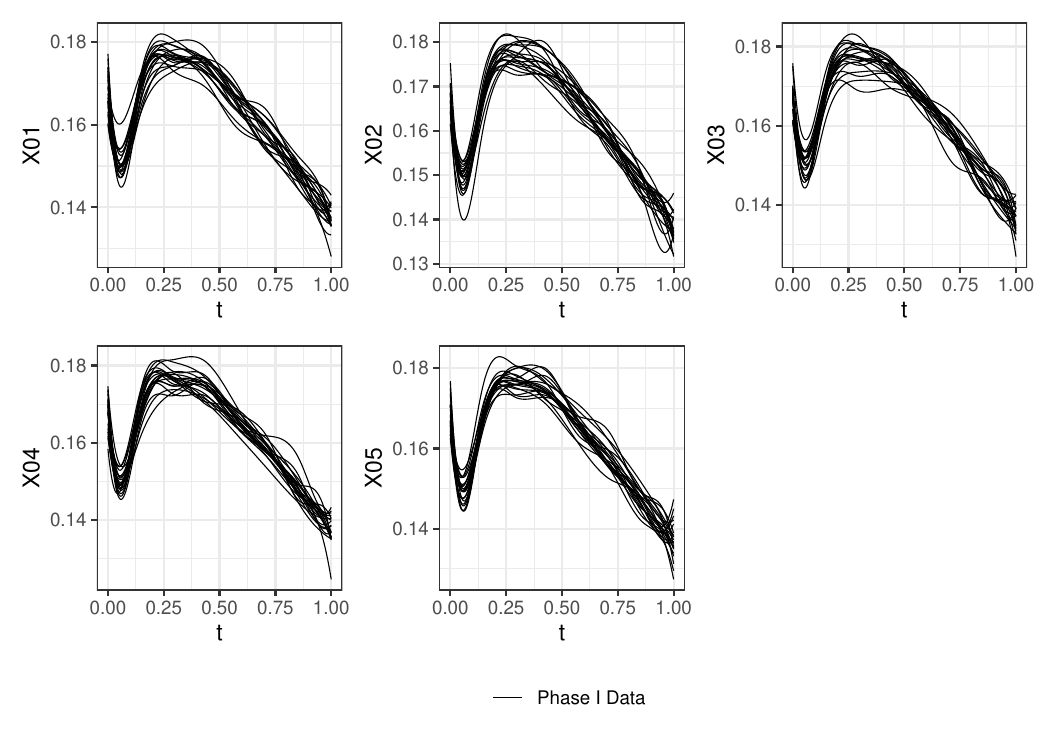}
    \caption{An example of 20 randomly generated IC observations}   
    \label{fig:PHASE I data simulated}
\end{figure}

\begin{figure}
    \centering
        \includegraphics[width=.9\columnwidth]{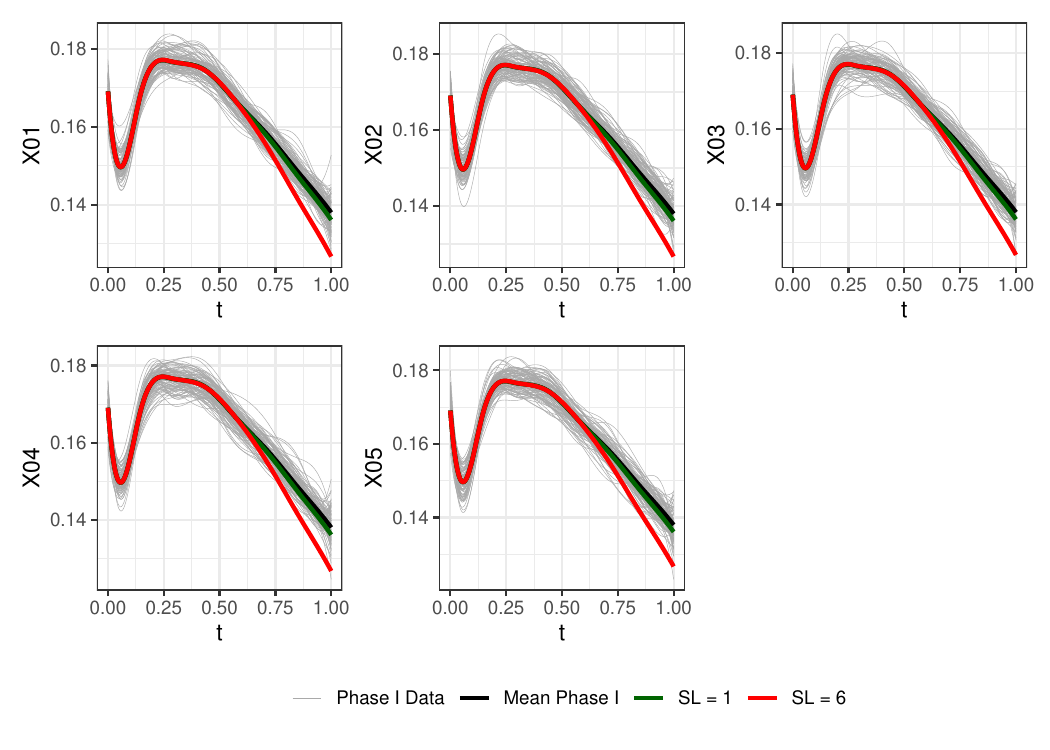}
    \caption{Simulated data in Scenario 1. Bold lines denote mean functions with $SL = 0$ (Mean Phase I), $SL=1$ and $SL=6$.}   
    \label{fig:Phase II data-Scenario 1}
\end{figure}

\begin{figure}
    \centering
        \includegraphics[width=.9\columnwidth]{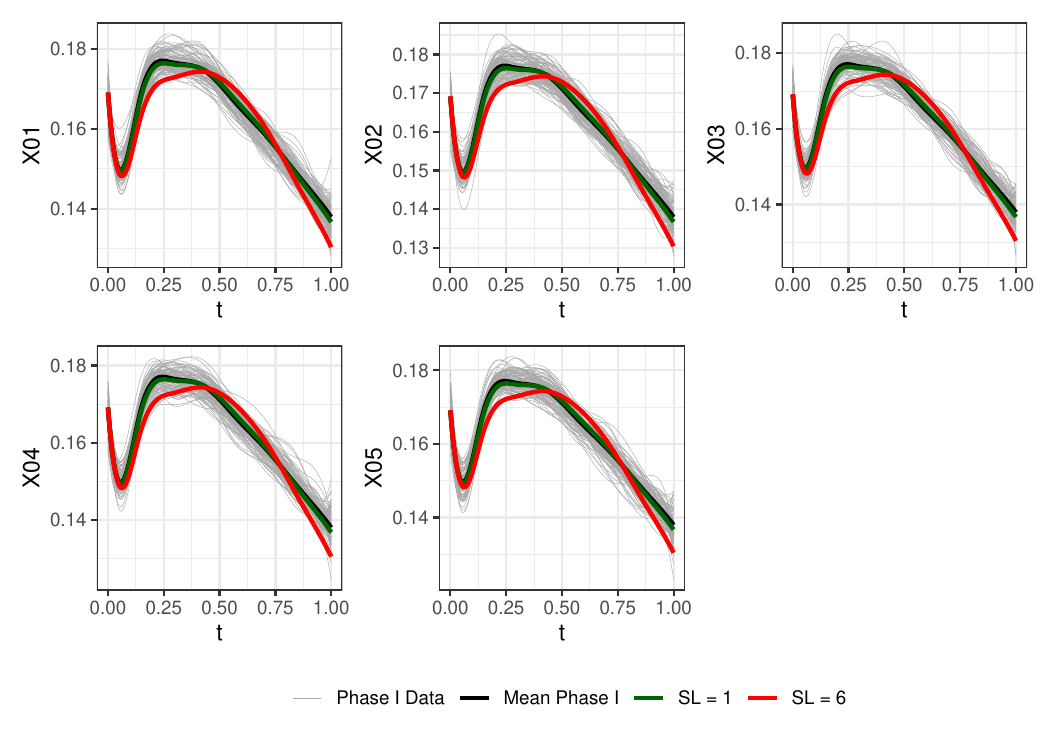}
    \caption{Simulated data in Scenario 2. Bold lines denote mean functions with $SL = 0$ (Mean Phase I), $SL=1$ and $SL=6$.}   
    \label{fig:Phase II data-Scenario 2}
\end{figure}

Then, the Phase I observations are generated through Equation \eqref{eq_modgen} with $B_E=B_P=0$. The Phase II sample is generated through Equation \eqref{eq_modgen} by considering the parameters listed in Table \ref{ta_2} for Scenario 1 and Scenario 2, with $\sigma_e=0.005$ and $\sigma=0.002$. In Scenario 1, OC observations  mimic a splash weld (expulsion) caused by excessive welding current, while, in Scenario 2, they are generated with a phase shift of the peak time caused by an increased force applied to the electrode used in the welding process \citep{xia2019online}.
\begin{table}
	\caption{Parameters used to generate the Phase II sample  for Scenario 1 and Scenario 2 and shift level $SL = \lbrace 0, 1, 2, 3, 4 , 5 , 6\rbrace$ in the simulation study.}
	\label{ta_2}
	\centering
	\resizebox{0.5\textwidth}{!}{
		\begin{tabular}{cM{0.1\textwidth}M{0.1\textwidth}M{0.1\textwidth}M{0.1\textwidth}}
			\toprule
			&\multicolumn{2}{c}{Scenario 1}&\multicolumn{2}{c}{Scenario 2}\\
			\cmidrule(lr){1-5}
		    &\multicolumn{2}{c}{\specialcell{$B_E=1$, $B_P=0$}}
			&\multicolumn{2}{c}{\specialcell{$B_E=0$, $B_P=1$}}
			\\
			\cmidrule(lr){2-3}\cmidrule(lr){4-5}
			$SL$&$M_E$&$M_P$&$M_E$&$M_P$\\
			\cmidrule(lr){2-3}\cmidrule(lr){4-5}
			    1&0.0019 & 0.00&0.00&0.025\\
		    2&0.0038 & 0.00&0.00&0.050\\
		    3&0.0056 & 0.00&0.00&0.075\\
		    4&0.0075 & 0.00&0.00&0.100\\
                5&0.0094 & 0.00&0.00&0.125\\
                6&0.0112 & 0.00&0.00&0.150\\
			\bottomrule
	\end{tabular}}
\end{table}
Finally, the generated data are assumed to be discretely observed at 25 equally spaced time points over the domain $\left[0,1\right]$.
For illustrative purposes, a sample of 20 randomly generated realizations of IC observations are shown in Figure \ref{fig:PHASE I data simulated}.
Whereas, Figure \ref{fig:Phase II data-Scenario 1}  and Figure \ref{fig:Phase II data-Scenario 2} show a sample of $100$ randomly generated observations in Scenario 1 for models M1, M2, and M3, each type of shift with severity level defined by $M_E$ and $M_P$ in Table \ref{ta_2}.

\bibliographystyle{chicago}
{\small
\bibliography{ref}}